\documentclass[11pt]{amsart}
\usepackage{geometry}                
\usepackage{graphicx}
\usepackage{amssymb}
\usepackage{epstopdf}
\newcommand{\bsym}[1]{\ensuremath{\boldsymbol{#1}}}
\renewcommand{\vec}[1]{\ensuremath{\mathbf{#1}}}
\newcommand{\set}[1]{\ensuremath{\mathcal{#1}}}

\newcommand{\Expect}{\ensuremath{\mathbb{E}}}

\newcommand{\mxm}{\ensuremath{\text{maximize}}}

\newcommand{\wrt}{\ensuremath{\mathrm{with \; respect \; to}}}
\newcommand{\sto}{\ensuremath{\mathrm{subject \; to}}}

	\usepackage{hyperref}
	\hypersetup{
	    bookmarks=true,         
	    unicode=false,          
	    pdftoolbar=true,        
	    pdfmenubar=true,        
	    pdffitwindow=false,     
	    pdfstartview={FitH},    
	    pdftitle={My title},    
	    pdfauthor={Author},     
	    pdfsubject={Subject},   
	    pdfcreator={Creator},   
	    pdfproducer={Producer}, 
	    pdfkeywords={keywords}, 
	    pdfnewwindow=true,      
	    colorlinks=true,       
	    linkcolor=red,          
	    citecolor=magenta,        
	    filecolor=cyan,      
	    urlcolor=blue           
	}

\title{Should Optimal Designers Worry About Consideration?}

\author{Minhua Long}
\address{Mechanical Engineering\\
	Iowa State University\\
	Ames, Iowa 50014}
\email{mhlong@iastate.edu}

\author{W. Ross Morrow}
\address{Ford Research and Innovation Center, Palo Alto\\
	Palo Alto, California 94304}
\email{wmorro13@ford.com}
\thanks{This research was supported by Iowa State University and NSF's ESD Program (\#1334764), and was primarily undertaken while W. Ross Morrow was an Assistant Professor of Mechanical Engineering at Iowa State University. Erin MacDonald at Stanford University, Phil Keenan at GM Market Research, and John Hauser at MIT's Sloan School of Business provided helpful suggestions. Erin MacDonald, Daria Dzyabura (NYU Stern School), and John Hauser graciously provided and assisted with use of the results reported in \cite{Urban2010, Dzyabura2011}. A previous version of this paper appears in the proceedings of ASME IDETC 2014, paper \#34493, and a full version is forthcoming in the Journal of Mechanical Design.}

\begin{document}

\maketitle


\begin{abstract}
{\it 
Consideration set formation using non-compensatory screening rules is a vital component of real purchasing decisions with decades of experimental validation. Marketers have recently developed statistical methods that can estimate quantitative choice models that include consideration set formation via non-compensatory screening rules. But is capturing consideration within models of choice important for design? This paper reports on a simulation study of a vehicle portfolio design when households screen over vehicle body style built to explore the importance of capturing consideration rules for optimal designers. We generate synthetic market share data, fit a variety of discrete choice models to the data, and then optimize design decisions using the estimated models. Model predictive power, design ``error'', and profitability relative to ideal profits are compared as the amount of market data available increases. We find that even when estimated compensatory models provide relatively good predictive accuracy, they can lead to sub-optimal design decisions when the population uses consideration behavior; convergence of compensatory models to non-compensatory behavior is likely to require unrealistic amounts of data; and modeling heterogeneity in non-compensatory screening is more valuable than heterogeneity in compensatory trade-offs. This supports the claim that designers should carefully identify consideration behaviors before optimizing product portfolios. We also find that higher model predictive power does not necessarily imply better design decisions; that is, different model forms can provide ``descriptive'' rather than ``predictive'' information that is useful for design.
}
\end{abstract}


\section{INTRODUCTION}

Conventional discrete choice models \cite{Ben-Akiva1988, Train2009} have been applied in design for market systems \cite{Wassenaar2003, Michalek2004, Michalek2005, Besharati2006, Shiau2009, Hoyle2010,  MacDonald2010, Michalek2011} in the past decade. Generally, the choice model serves to forecast demand as a function of product features, thus enabling design decisions that maximize forecast profits. These conventional choice models share the assumption that individuals choose by processing and weighing all attributes, for all alternatives, when maximizing utility. According to this assumption choice is a compensatory decision making process where tradeoffs can take place across all features and all alternatives: in particular, shortcomings in one attribute can always be compensated by making others sufficiently attractive. Empirical studies have shown the opposite: people often use ``fast and frugal'' non-compensatory rules to eliminate options when faced with task complexity \cite{Payne1976}, time pressure \cite{Rieskamp2008}, information cost \cite{Broder2000} and memory requirements \cite{Broder2003}. The use of such heuristics$-$decision rules that ignore information$-$is widespread and beneficial \cite{Gigerenzer2011}. This paper investigates the importance of including consideration behavior when making design decisions. 

The awareness of the use of non-compensatory rules among consumers has changed the traditional concept of the choice set in choice modeling \cite{Shocker1991}. Instead of assuming only a universal choice set with all alternatives, \textit{consideration-sets} \cite{Roberts1997, Hauser2009} have become a topic of active research. Consideration-sets are subsets of the universal set that are chosen by individuals following internal, non-compensatory rules. Building on early research on non-compensatory decision models \cite{Dawes1964, Einhorn1970, Tvesky1972}, non-compensatory rules proposed for consideration set formation include conjunctive, disjunctive, subset conjunctive, and even lexicographic rules; see \cite{Hauser2014} for further background and examples. Accepting consideration implies that identification of the structure and distribution of screening rules is an important empirical task that much recent research addresses, as reviewed in Section \ref{SEC: Review_CTC}. 

But is modeling consideration important when making design decisions? Marketers have only shown the advantage of modeling consideration through improvement in model predictive accuracy, though this has been accomplished across a wide variety of product categories including cameras, batteries, automobiles, cellphones, and computers \cite{Gilbride2004, Jedidi2005, Hauser2009, Ding2011,Yee2007}; e.g., see Table \ref{TBL: summaryCTC} below. Simulation experiments have illustrated the limits of classical compensatory models including the multinomial and random coefficient (Mixed) Logit models when modeling non-compensatory choice behavior \cite{Johnson1984, Andrews1998, Andrews2008}. Existing engineering studies demonstrate how design can include consideration in choice model structure, and how this might affect decisions \cite{Besharati2006a, Morrow2014b}, but have not compared the performance of compensatory and non-compensatory models when both types of models are estimated on the same data with a comparable level of system knowledge. Even if compensatory models do not represent non-compensatory choice behavior well, could they still suggest product designs similar to designs that are optimal for true, non-compensatory behavior? If the non-compensatory behavior is modeled directly, how much closer could a firm get to true optimal designs? What is the difference of the \textit{value} of the chosen designs, e.g. profits, between designs chosen using compensatory versus non-compensatory models?

We describe a simulation study that examines how well compensatory models perform in 1) recovering non-compensatory choice behavior, 2) suggesting design decisions near to ideal optimal decisions, and 3) suggesting designs that capture all potential profitability. Our ``synthetic data'' \cite{Dzyabura2011} simulation experiment has the following steps: 
\begin{description}
	\item[1:] Define a synthetic population with known ``true'' choice behavior;
	\item[2:] Simulate responses of this population to a sequence of ``markets'' with randomly generated product profiles;
	\item[3:] Estimate compensatory and non-compensatory models from the responses and validate predictive power;
	\item[4:] Optimize design decisions with the estimated models and evaluate design profit using the ``true'' behavior.
\end{description}
Synthetic data experiments are an effective method for detecting choice model properties in specific situations or when testing the validity of an estimation approach \cite{Andrews2008, Kohli2007, Kropko2011, Dzyabura2011}. We extend this paradigm to also include the quality and value of decisions made using estimated models, the ultimate goal of choice modeling within engineering design. The synthetic data experiment allows us to measure the divergence of design decisions and outcomes from ideal values that can be obtained only by knowing the true behavioral model. We describe an ``econometric-style'' (revealed preference) experiment that uses aggregate share data to estimate choice models. An alternative perspective, more common in marketing, samples the population for respondents to choice and/or consideration-based conjoint surveys (stated preference). Both perspectives have value, as is discussed in \cite{Train2009}, pg.152. Both types of models have also been used in design \cite{Michalek2004, Wassenaar2005}. 

Several observations are enabled by the experiment. As would be expected, modeling consideration with a non-compensatory model results in the best design and pricing decisions when the population exhibits matching non-compensatory behavior. Conventional compensatory models can reasonably support profitable design decisions, however, with several caveats: conventional models might require more data than is reasonably available to capture non-compensatory behaviors, can suggest simplistic product portfolios, can be sensitive to sample variance in the training data, and don't forecast the value of design decisions well even if those decisions couldn't be improved with a better model. Overall, modeling heterogeneity in the screening rules used to form consideration sets captures more value to design than modeling heterogeneity in the compensatory stage. A similar observation has been made by Andrews et al. \cite{Andrews2008}. Finally, while assuming that better model predictive power implies better design decisions is reasonable, it is not necessarily true: models with {\em lower} predictive power can suggest {\em more} profitable designs. We hope our case study will motivate market systems researchers to further examine what consideration behaviors exist in their product categories and how these behaviors might influence optimality of chosen designs.


The rest of paper is organized as follows: Section \ref{SEC: Review_CTC} reviews the consider-then-choose model construction and estimation studied in marketing research. Section \ref{SEC:CaseStudy} describes the simulation framework and synthetic data generation process. Section \ref{SEC:Estimation} and \ref{SEC:Design_optimization} respectively provides details of model estimation and design optimization. Section \ref{SEC:Results} presents our results, followed by discussion in Section \ref{SEC:Discussion}. Section \ref{SEC:Conclusion} concludes.


\section{CONSIDER-THEN-CHOOSE MODELS}
\label{SEC: Review_CTC}

A consider-then-choose model can be described as follows.  Suppose the universal choice set is $\set{J}=\{1,..., J\}$. A {\em consideration set} indexed by $r=1, ..., R$, denoted as $\set{C}_r \subset \{ 1 ,\dotsc , J \}$ is defined by a set of {\em screening rules} $\vec{s}_r = [ s_{r,1}, ..., s_{r,L_r} ]$. For conjunctive rules, $\set{C}_r$ can be written as:
\begin{equation}
	\label{EQN:ConsiderSet_conjunctive}
	\set{C}_r(\vec{X},\vec{p})
		= \Big\{ \; j \in \{1,\dotsc,J\} \; : \; \vec{s}_r(\vec{x}_j,p_j) \leq \vec{0} \; \Big\}
\end{equation}
The screening rules depend on product features $\vec{x}_j$, price $p_j$ as well as other rule-specific parameters. This definition means that a product needs to satisfy {\em all} the screening rules to be a member in the corresponding consideration set. For example, the consideration set 
\begin{align*} 
	\set{C}_r(\vec{X},\vec{p}) = &\{ \text{all vehicles $j$ with price $p_j$ under \$20,000} \\
	&\quad \text{AND fuel economy $e_j$ over 30 mpg} \} 
\end{align*}
can be defined by 
\begin{equation*}
	\vec{s}_r(\vec{x}_j,p_j) 
		= \begin{pmatrix} p_j - 20,000 \\ 30 - e_j \end{pmatrix} \leq \vec{0}.
\end{equation*} 
This structure is consistent with the forms used in the marketing literature, although marketers often define screening rules in terms of indicators instead of inequalities. See, for example \cite{Gilbride2004, Liu2011, Dzyabura2011}. These representations can be transformed into the structure presented here. 

Given a collection of screening rules and the associated consideration set, let the conditional probability that product $j$ is chosen within the set be $P_{j|\set{C}_r}$ and let the probability that the consideration set $\set{C}_r$ is formed be $P_{\set{C}_r}$. Then the choice probability $P_j$ can be written as a weighted sum of the choice probabilities across all possible consideration sets:
\begin{equation}
	\label{EQN:CTCjointProb}
	P_j = \sum_r P_{j|\set{C}_r} P_{\set{C}_r}
\end{equation}
Hauser \cite{Hauser2014} calls such models ``consideration'' or ``choice set explosion'' models, as they are subject to combinatorial explosion in the number of parameters needed to capture consideration set occurrence. Empirical methods estimate $P_{\set{C}_r}$ directly, rather than uncovering structure behind screening by identifying the rules $\vec{s}_r$. Manrai and Andrews \cite{Manrai1998} provide a thorough review of studies applying Eqn. (\ref{EQN:CTCjointProb}) to scanner panel data. Note that Eqn. (\ref{EQN:CTCjointProb}) can also be considered a type of random coefficients (Mixed) Logit model, though not one with {\em normally} distributed coefficients. This structure has also been found to be similar to a nested Logit model, as we detail in Sec. \ref{SUBSEC:NestedLogitCTC} below. 

Preference-conditional choice probabilities then take the following form: 
\begin{equation}
	\label{EQN:CTCProb}
	P_{j|\set{C}_r}(\vec{X},\vec{p}\mid\bsym{\theta}) 
		= \left\{\begin{aligned}
			&\frac{e^{u(\vec{x}_j,p_j, \bsym{\theta})}}{1 + \sum_{k \in \set{C}_r} e^{u(\vec{x}_k,p_k,\bsym{\theta})}}
				&&\text{if } j \in \set{C}_r \\
			&\quad\quad\quad\quad 0
				&&\text{if } j \notin \set{C}_r
		\end{aligned} \right .
\end{equation}
where utility $u(.)$ is a function of product characteristics $\vec{x}_j$ and price $p_j$ given coefficients $\bsym{\theta}$ that measure preferences. The utility coefficients $\bsym{\theta}$ can be assumed to be homogeneous across the population or take a random coefficients form to include heterogeneity (which requires a Monte-Carlo integral of the simple Logit form above). This formula can, in principle, be extended to capture heterogeneity across consideration sets by allowing a nontrivial joint distribution between coefficients $\bsym{\theta}$ and consideration sets.

Methods used in early studies to discover non-compensatory screening rules included tracing and protocol analysis \cite{Payne1976, Bettman1980} in which respondents' decision processes were self-reported or tracked. Shortcomings of this type of method have been reported and include inconsistencies between the stated screening criteria and observed choices from the same individual \cite{Green1988}. More recent research shows that the accuracy of direct elicitation approaches can be improved by designing experiments that are incentive-compatible: for example, by participating in a survey in which respondents describe their screening rules for new vehicles, they have a decent chance of actually winning a vehicle described \cite{Ding2011}. Estimation tools that are widely applied in traditional discrete choice analysis, e.g. maximum likelihood and Bayesian methods, can also be used in non-compensatory model parameter estimation \cite{Jedidi1996, Jedidi2005, Gilbride2004}. These methods may, however, suffer from high computation costs due to exponential growth in the number of possible consideration sets as the number of attributes and/or attribute levels grows. Machine learning techniques have recently been adapted to circumvent this problem by applying greedoid methods \cite{Yee2007, Kohli2007} or low-dimensional parameterizations of screening rule likelihood \cite{Urban2010, Dzyabura2011}. Broadly speaking, marketing research has demonstrated predictive power improvement by modeling consideration; see Table \ref{TBL: summaryCTC}. 

\begin{table*}
\caption{Recent consider-then-choose models constructed from stated preference data, compared to compensatory models estimated in the same study. Abbreviations: HB - Hierachical Bayes; MLE - Maximum Likelihood; HR - ``hit rate'' (frequency of correct prediction on hold-out samples); KLD - Kullback-Liebler Divergence; TAU - Kendall's Tau \cite{Hauser1978}. }
\label{TBL: summaryCTC}
\centering
\begin{footnotesize}
\begin{tabular}{clllccc}
 &  &  &  &  \multicolumn{3}{c}{\% Improvement} \\
Reference & Product & Compensatory Model &  Consider-then-Choose Model & HR & KLD & TAU \\
\hline
\cite{Kohli2007}	&Laptops  	& LP Logit			& Greedy, Lexicographic			& & & 0\% 		\\
\cite{Jedidi2005}	&Batteries 	& MLE Logit		& MLE, Subset conjunctive		& 1.1\%     		\\
\cite{Hauser2009}	& Cameras 	& HB Logit		& HB, Conjunctive screening		& 7.1\%      		\\
\cite{Yee2007}		&Smartphones &HB Ranked Logit 	& Lexicographic by aspects		& 8.7\%  		  	\\
\cite{Ding2011}		&Cellphones	&HB Logit			& Unstructured Direct Elicitation	& & 9.1\%   		\\
\cite{Swait2001}	&Rental Cars    &MLE Logit		& ``Cut-off rules'' (conjunctions)		& & 14.0\%$^{\text{(a)}}$ 	\\
\cite{Hauser2010}	&GPS Units	& HB Logit 		& Greedy, Lexicographic			& 4.5\% & 54.5\%  \\
\cite{Dzyabura2011}	&Vehicles		&HB Logit                 & Adaptive question HB			& 44.1\% & 16.7\% \\
\hline
\multicolumn{7}{l}{(a) \cite{Swait2001} characterized improvement with improvement in log-likelihood, which is proportional to KLD.}
\end{tabular}
\end{footnotesize}
\end{table*}

In principle the specification reviewed above can be estimated from choice data with classical tools such as Maximum Likelihood Estimation (MLE) and Bayesian methods. To facilitate this, the representation of consideration set probability $P_{\set{C}_r}$ has taken different forms. Swait \cite{Swait1987} introduced a random component into the screening rules so that with an assumed distribution $P_{\set{C}_r}$ can be derived based on the probability any alternative is acceptable. Ben-Akiva \cite{Ben-Akiva1995} extended this random consideration set generation model by specifying the availability probability as Logit form. Instead of defining $P_{\set{C}_r}$ through parameterized screening rules, Chiang et al. \cite{Chiang1999} assumed consideration set probabilities have a Dirichlet distribution across the population. Gilbride and Allenby \cite{Gilbride2004} avoided the enumeration of consideration sets by using a reduced form choice probability and Markov Chain Monte-Carlo methods to sample from the posterior distribution of the allowable screening criteria values. Exponential growth in the number of possible consideration sets and rules makes consider-then-choose models difficult to estimate in practice. This challenge motivated researchers to develop methods that apply to ``consider'' stage observations to infer screening rules with more attributes and complexity. For example, MLE methods have been used on ``acceptable/unacceptable'' responses to the profiles to estimate the probability that a particular attribute level is acceptable \cite{Jedidi2005}. Dzyabura \& Hauser \cite{Dzyabura2011} model a case where capturing the distribution of screening rules would require $2^{53}$ parameters, too many for a direct estimation strategy. They develop an adaptive question survey strategy to estimate conjunctive screening rules by parameterizing screening rule likelihood presuming feature acceptability is independent, obtaining a model with only 53 parameters per respondent.   


\section{CASE STUDY: VEHICLE DESIGN UNDER BODY STYLE SCREENING}
\label{SEC:CaseStudy}

We simulate a stylized model of the new vehicle market with potential purchasers that screen over vehicle body style. Empirical studies have shown body style screening in both self-reported surveys \cite{Punj2002} and statistical inferences \cite{Dzyabura2011}. Body style also significantly impacts the engineering relationships between other features in vehicle design. We often refer to the synthetic behavior described below as the ``true'' behavior. We do not use this terminology to suggest this is how households actually choose new vehicles. This is only a shorthand appropriate for the context of the simulation experiment. 


\subsection{Synthetic Behavior}
\label{SUBSEC:SyntheticBehavior}

Our population is a mix of groups that screen over $B = 9$ vehicle body styles listed in Table \ref{TAB:AggregatedAcceptabilitiesBS}. Vehicles are described by  fuel economy ($e$), acceleration ($a$), price ($p$) and a $B$-element binary vector $\bsym{\delta}$ for which $\delta_b = 1$ if, and only if, the vehicle has body style $b$ (thus $\sum_b \delta_b = 1$). Let $\vec{s} \neq \vec{0}$ be a $B$-element binary vector defining which body styles are ``acceptable'' to a given individual in the population; we refer to these vectors succinctly as ``screening rules.'' Unlike $\bsym{\delta}$, which can have only one element equal to 1, $\vec{s}$ can have any number of elements equal to 1. An individual with screening rule $\vec{s}$ considers only those vehicles with body styles $b$ such that $s_b = 1$ or, equivalently, $\vec{s}^\top\bsym{\delta} \geq 1$. In the notation of Eqn. (\ref{EQN:ConsiderSet_conjunctive},\ref{EQN:CTCProb}) we can index individuals by screening rules $\vec{s}$ and define
\begin{equation}
	\set{C}_{\vec{s}}(\triangle) = \Big\{ j \in \{1,\dotsc,J\} : 1 - \vec{s}^\top\bsym{\delta}_j \leq 0 \Big\}
\end{equation}
where $\triangle = ( \bsym{\delta}_1,\dotsc,\bsym{\delta}_J)$ is a matrix of binary body style vectors. 

The fraction of individuals in the population with a particular screening rule $\vec{s}$ is given by a probability mass function $\alpha(\vec{s})$ drawn from the results of the empirical study reported in \cite{Urban2010, Dzyabura2011}. This study estimated conjunctive screening rules with a Bayesian adaptive question method for 874 respondents. More specifically, we take $\alpha$ to be the empirical frequency distribution (over respondents) of the modal (most probable) rules $\vec{s}$ for the posterior distribution. Out of the 874 respondents, 219 distinct most-probable conjunctive screening rules were estimated, and every body style is acceptable to {\em some} individual. See Table \ref{TAB:AggregatedAcceptabilitiesBS} for the aggregated acceptability of different body styles over the full respondent pool.

\begin{table}
	\begin{center}
	\begin{footnotesize}
	\caption{Percentage of respondents accepting the given body style, as reported in \cite{Dzyabura2011}.}
	\label{TAB:AggregatedAcceptabilitiesBS}
	\begin{tabular}{cccccccccc}
	\rotatebox{90}{Sports car} 
		& \rotatebox{90}{Hatchback} 
		& \rotatebox{90}{Compact sedan} 
		& \rotatebox{90}{Standard sedan} 
		& \rotatebox{90}{Crossover} 
		& \rotatebox{90}{Small SUV} 
		& \rotatebox{90}{Full-size SUV} 
		& \rotatebox{90}{Pickup truck} 
		& \rotatebox{90}{Minivan} \\ \hline
	16\%
		& 19\%
		& 38\%
		& 42\%
		& 38\%
		& 39\%
		& 29\%
		& 18\%
		& 10\% \\ \hline
	\end{tabular}
	\end{footnotesize}
	\end{center}
\end{table}

Conditional on using a screening rule $\vec{s}$, individuals choose from those vehicles in $\set{C}_{\vec{s}}(\triangle)$ by maximizing the random utility:
\begin{align}
	\label{EQN:TrueRandomUntility}
	 U_j &= \hat{u}(e_j,a_j,p_j;\hat{\bsym{\theta}} ) + \set{E}_j \\
	\label{EQN:TrueRepresentiveUtility}
		\hat{u}(e,a,p;\hat{\bsym{\theta}}) &= - \exp\{\hat{\theta}_p\} p 
					+ \frac{\hat{\theta}_e}{e}
					+ \frac{\hat{\theta}_a}{a}
					+ \hat{\theta}_0
\end{align}
for random coefficients $\hat{\theta}_l \sim \set{N}(\hat{\mu}_l,\hat{\sigma}_l)$ ($l = p,e,a,0$). $\set{N}(\hat{\mu},\hat{\sigma})$ is the normal distribution with mean $\hat{\mu}$ and variance $\hat{\sigma}$. The exponential in the price coefficient ensures that lower prices are preferred, all other things being equal. The errors $\bsym{\set{E}} = (\set{E}_0,\set{E}_1,\dotsc,\set{E}_J)$ are i.i.d. extreme value variables mean-shifted towards zero. The resulting screening-conditional sub-populations thus follow a Mixed Logit model. There are no correlations between screening rules and preference over vehicle attributes and price, but there is heterogeneity in the population. 

\begin{table}
	\begin{center}
	\begin{footnotesize}
		\caption{Means and Variances of random coefficients ($\hat{\theta}$'s) in synthetic population utility function, Eqn. (\ref{EQN:TrueRandomUntility}). $\set{N}(\hat{\mu},\hat{\sigma})$ refers to a normally distributed variable with mean $\hat{\mu}$ and variance $\hat{\sigma}^2$. Values based on the model from \cite{Whitefoot2013}.}
		\label{TAB:TrueCoeffs}
		\begin{tabular}{llcrr}
			\\
							&		&		& \multicolumn{2}{c}{Random Coefficient} \\
			Attribute			&  	& Utility 	& Mean ($\hat{\mu}$) & Variance ($\hat{\sigma}$) \\ \hline
			Price				& ($p$)	& $-\exp\{\set{N}(\hat{\mu},\hat{\sigma})\}p$ 	
									& $2.0$ 		& $0.1$ \\
			Fuel Economy 		& ($a$)	& $\set{N}(\hat{\mu},\hat{\sigma})/e$ 		
									& $-36.8$ 	& $2.2$  \\		
			Acceleration 		& ($e$)	& $\set{N}(\hat{\mu},\hat{\sigma})/a$ 		
									& $ 11.3$ 		& $0.3$ \\		
			Constant			& ($-$) 	& $\set{N}(\hat{\mu},\hat{\sigma})$		
									& $-23.2$ 		& $0.5$
						\\ \hline
		\end{tabular}
	\end{footnotesize}
	\end{center}
\end{table}


\subsection{Market Share Simulation}
\label{SUBSEC:MarketShareSimulation}

We simulate sales data set that might be collected from multiple, separate new vehicle markets. The vehicles in each market form a ``universal'' choice set for the consumers in that market. A data set for estimation then consists of vehicle market shares in $M$ separate markets indexed by $m$. For each market, we draw a set of $J_m$ vehicles, denoted $\set{J}_m$. The profile of vehicle $j$ in market $m$ is given by drawing fuel economy ($e_{j,m}$), acceleration ($a_{j,m}$), price ($p_{j,m}$) and body style ($\vec{b}_{j,m}$) from a uniform distribution respectively on intervals $[5,50]$ (mpg), $[2,15]$ (s) and $[10,60]$ (10k\$) and $\{1,\dotsc,B\}$. An alternative consistent with our optimal design problem presented below would be to draw sets of vehicles that satisfy our assumed design constraints. This is possible, and better matches the stylized market modeling paradigm we employ. However random draws are likely to give us better information about choice behavior than correlated draws, and thus allow us to focus more completely on choice model quality. To investigate statistical properties of model estimation and use with stochastic data generation and choice outcomes, this process is repeated with different random seeds.

Given product profiles in market $m$, we draw $N_m$ choice observations in which individuals can purchase one of the vehicles or choose not to purchase any vehicle (choose the ``outside good''). $N_m$ individuals are drawn from the synthetic population by drawing $N_m$ screening rules $\vec{s}_i$ from the distribution $\alpha(\vec{s})$ along with associated random coefficients $\bsym{\theta}_i$. Shares $S_{j,m}$ for each vehicle $j$ in each market $m$ (and the outside good) are then generated by maximizing random utilities (utilities plus error term) foreach individual over their consideration set. See \cite{Long14} for an explicit algorithm. 


\section{CHOICE MODELS}
\label{SEC:Estimation}

We examine four choice model specifications: Multinomial Logit (MNL), Random Coefficients Logit (RCL), Nested Multinomial Logit (NML) and Consider-Then-Choose Logit (CTC) models. We assume that all models incorporate the prior information that body style plays a role in consumer decision, but different model specifications incorporate this piece of information in different structures: MNL and RCL model assume the tradeoffs between body style and other attributes, NML uses nests that separate body styles, thus constructing a two-stage but yet compensatory process; CTC models the frequency of any possible consideration sets, based on body style, along with compensatory choices conditional on consideration set. Note that the true behavior of the synthetic population exhibits characteristics of both non-compensatory screening and heterogeneity in compensatory stage. Thus all the models are misspecified on at least one behavioral feature. The comparison between these models will thus illustrate the consequence of failing to capture different behavioral features.

Coefficients in all models are estimated by maximizing the log-likelihood with respect to the coefficients \cite{Train2009}. For a general choice model with probabilities $P_{j,m}(\bsym{\theta})$ for coefficients $\bsym{\theta}$, this takes the form:
\begin{equation}
	\begin{aligned}
		\mxm &\quad \sum_{m=1}^M \sum_{j\in \mathcal{J}_m \cup \{0\}} S_{j,m} \log \Big( P_{j,m}(\bsym{\theta}) \Big) \\
		\wrt	&\quad \bsym{\theta} \text{ (plus possible constraints)}\\
	\end{aligned}
\end{equation}
We abbreviate this process by ``MLE'' and, for brevity, do not explicitly list each MLE problem below. Instead we define the choice probability model and list any constraints imposed on the coefficients as this is sufficient to recreate our process. 

\subsection{Multinomial Logit Model (MNL)}

The MNL model takes the utility of product $j$ to be
\begin{equation}
\label{EQN:Logit Utility}
	\begin{aligned}
		&u_{j,m}^{MNL}( \bsym{\theta}) 
			=  - \exp\{\theta_p\} p_{j,m} + \frac{\theta_e}{e_{j,m}} + \frac{\theta_a}{a_{j,m}}
											    + \sum_{b=1}^B \theta_b \delta_{j,m,b} +\theta_0
	\end{aligned}
\end{equation}
giving choice probabilities
\begin{equation}
\label{EQN:Logit Probability}
		P^{MNL}_{j,m}(\bsym{\theta}) 
			= \frac{\exp \{ u_{j,m}^{MNL}( \bsym{\theta}) \} }
				{1 + \sum_{k\in \set{J}_{m}}  \exp \{ u_{k,m}^{MNL}( \bsym{\theta}) \} }
\end{equation}
As with the true behavior, the ``$\exp$'' term in the price coefficient ensures that the price coefficient is negative, and thus lower prices are preferred (all other attributes being equal). The ``no buy'' or outside good probability is $P^{MNL}_{0,m}(\bsym{\theta}) = 1 - \sum_{j\in\set{J}_m} P^{MNL}_{j,m}(\bsym{\theta})$. The body style coefficients are not independently identified from the constant $\theta_0$ because body styles are represented by dummies that always sum to one. We use ``effects coding'' \cite{Bech2005} and constrain the coefficients over body styles to sum to zero. We could, equivalently, leave the constant term or one of the body style dummies out of the specification. We prefer to include the constant and all dummies with effects coding as this allows us to capture only body style specific variations in utility with the coefficients on the body style dummies. 


\subsection{Random Coefficients Logit Model (RCL)}
\label{SUBSEC:RC}

In the RCL model, choice probabilities $P_{j,m}^{RCL}$ are defined for each vehicle $j$ in each market $m$ by
\begin{equation}
	\label{EQN:mixed_prob}
	\begin{aligned}
		P_{j,m}^{RCL}(\bsym{\mu},\bsym{\sigma})
			= \int P_{j,m}^{MNL}(\bsym{\theta}) \phi(\bsym{\theta}\mid\bsym{\mu},\bsym{\sigma})d\bsym{\theta}
	\end{aligned}
\end{equation}
for $P_{j,m}^{MNL}$ as given in Eqn. (\ref{EQN:Logit Probability}). All random coefficients as written in Eqns. (\ref{EQN:Logit Probability}-\ref{EQN:mixed_prob}) are assumed to be normally distributed, $\theta_l = \mathcal{N}(\mu_l,\sigma_l)$ $l = p, a, e, 1, ..., B$, with mean $\mu_I$ and variance $\sigma_I^2$. Note, however, that this implies that the price coefficient will be log-normal (e.g., \cite{Boyd1980, Berry2004}). The density $\phi$ is thus a product of $4+B$ independent normal densities each having two parameters. The RCL model thus has $8 + 2B$ coefficients we must estimate, two of which enter into the utility function nonlinearly. 

Given synthetic revealed preference data we estimate the parameters $(\bsym{\mu},\bsym{\sigma})$ using simulated MLE \cite{Train2009}. We perform Monte-Carlo sampling over random coefficients to obtain $I$ samples $\bsym{\theta}_i \sim \set{N}(\bsym{\mu},\mathrm{diag}(\bsym{\sigma}^2))$ and simulated RCL choice probabilities
\begin{equation}
	\tilde{P}_{j,m}^{RCL}(\bsym{\mu},\bsym{\sigma})
			  = \left(\frac{1}{I}\right)
			  		\sum_{i=1}^I P_{j,m}^{MNL}(\bsym{\theta}_i). 
\end{equation} 
We use $I = 1,000$ Monte-Carlo samples throughout this study if not otherwise mentioned. Similar to the Logit model estimation, the perfect correlation between body style coefficients can lead to multiple estimators that give the same choice probability. Therefore the mean coefficients on body styles are also constrained so that their sum equals to zero. Note that this does {\em not} imply that $\sum_b \theta_b = 0$ with probability one, but only that $\Expect[ \sum_b \theta_b ] = 0$. 


\subsection{Nested Multinomial Logit Model (NML)}

We also examine a NML model in which vehicles with the same body style are assigned to the same nest. Suppose product $j$ in market $m$ belongs to nest $\set{N}_{b(j),m}$ where $b(j)$ is the body style of product $j$. The probability product $j$ is chosen in market $m$ is
\begin{equation}
\label{EQN:Nested Logit Probability}
	P^{NML}_{j,m} (\bsym{\theta}) = P_{j \mid b(j),m} P_{b(j),m}
\end{equation}
where $P_{b,m}^N$ is the probability that any product from nest $\set{N}_b$ is chosen in market $m$ and $P_{j|b(j),m}^C$ is the probability that product $j$ is chosen in market $m$, conditional on nest $b(j)$ being chosen. $P_{j \mid b(j),m}^C$ follows the logit formula in which only non-body style features are involved in the utility: 
\begin{equation}
\label{EQN:Conditional Choice Probability within Nest}
	P_{j\mid b(j),m}^C(\bsym{\theta}) 
		= \frac{\exp\{u_{j,m}^{NML}(\theta_p,\theta_e,\theta_p)\}}{\sum_{k\in \set{N}_{b(j),m}}\exp\{u_{k,m}^{NML}(\theta_p,\theta_e,\theta_a)\}}
\end{equation}
with utility within the nest defined as:
\begin{equation}
	u_{j,m}^{NML}(\theta_p,\theta_e,\theta_a) 
		= - \exp\{\theta_p\} p_{j,m} + \frac{\theta_e}{e_{j,m}} + \frac{\theta_a}{a_{j,m}}
\end{equation}
The choice of nest depends on the ``nest utility''
\begin{equation}
\label{EQN:Nest Utility}
\begin{aligned}
	V_{b,m}(\theta_p,\theta_e,\theta_a) 
		=  \log \left( \sum_{j\in \set{N}_{b,m}} \exp\{ u_{j,m}^{NML}(\theta_p,\theta_e,\theta_a) \} \right)
\end{aligned}
\end{equation}
and also takes the logit form
\begin{equation}
\label{EQN:Nest Probability}
P_{b,m}^N(\bsym{\theta},\bsym{\lambda}) 
	= \frac{\exp\{ \theta_0 + \theta_b  + \lambda_b V_{b,m}(\theta_p,\theta_e,\theta_a) \}}
					{1 + \sum_{c=1}^B \exp\{ \theta_0 + \theta_c  + \lambda_c V_{c,m}(\theta_p,\theta_e,\theta_a) \}}
\end{equation}
We again constraint the body style dummies coefficient to sum to zero, for the same reason as in MNL and RCL models.

This formulation follows Daly's version of the NML \cite{Daly1987}, rather than the ``Generalized Extreme Value'' formulation given by McFadden \cite{McFadden1978}. The difference between two formulations is that McFadden's model uses 
\begin{equation*}
\begin{aligned}
	\theta_0 + \theta_b  + \lambda_b V_{b,m}\left(\frac{\theta_p}{\lambda_b},\frac{\theta_e}{\lambda_b},\frac{\theta_a}{\lambda_b}\right)
\end{aligned}
\end{equation*}
as the utility in Eqn. (\ref{EQN:Nest Probability}). This change is required for consistency with random utility maximization, but there is still debate about whether that is essential in the model \cite{Train2009}. Both versions have similarities with consideration behavior, as discussed below. 

\subsection{Consider-Then-Choose Logit Model (CTC)}
\label{SUBSEC:CTCLogit}

In the CTC model body styles are screened in the non-compensatory stage and do not enter the compensatory stage. Preference in compensatory stage is assumed to be homogeneous both among the population and across all consideration sets. The body styles screening rules are $\vec{S} = (\vec{s}_1, ..., \vec{s}_R)$ characterizing all $R = 2^B - 1$ possible consideration sets $\set{C}_1, ..., \set{C}_R$, except the ``null set'' in which no body style is considered. Each screening rule is coded as $B$-element binary vector $\vec{s}_r = (s_{r,1}, ..., s_{r,B})$ where $s_{r,b} = 1$ if body style $b$ is acceptable, $s_{r,b} = 0$ otherwise. Thus
\begin{equation}
	\set{C}_{r,m} = \{ j : \vec{s}_r^\top\bsym{\delta}_{j,m} \geq 1 \} 
\end{equation}
which means that a product will be considered as long as its body style is acceptable. 

The choice probability for product $j$ in market $m$ is
\begin{equation}
\label{EQN:CTCLogit Probability}
	P^{CTC}_{j,m} (\bsym{\theta},\bsym{\alpha})   
		= \sum_{r=1}^R \alpha_r 
				\left( \frac{ \exp\{u_{j,m}^{CTC}(\bsym{\theta})\} } 
					{ 1 + \sum_{k\in \set{C}_{r,m}} \exp\{u_{k,m}^{CTC}(\bsym{\theta})\} } \right)
\end{equation}
if $j \in \set{C}_{r,m}$ and zero otherwise, where $\alpha_r = \alpha(\vec{s}_r) $ is an estimator of the probability that a randomly drawn individual in the population has screening rule $\vec{s}_r$ and utilities are defined by:
\begin{equation}
\label{EQN:CTCLogit Utility}
	u_{j,m}^{CTC}(\bsym{\theta}) = - \exp\{\theta_p\} p_{j,m} + \frac{\theta_e}{e_{j,m}} + \frac{\theta_a}{a_{j,m}} +\theta_0
\end{equation}
We estimate this model with MLE, solving for both $\bsym{\theta}$ and $\bsym{\alpha} \in [\vec{0},\vec{1}]$, $\sum_r \alpha_r = 1$. These constraints are required to ensure that $\bsym{\alpha}$ is a probability mass function. 

Note that we directly estimate consideration set probability $P_{\set{C}_r} = \alpha(\vec{s}_r)$ rather than estimating parameters of the distribution $\alpha(\vec{s})$. Our case study is small enough to enable us to enumerate the consideration sets, requiring only $R = 511$ $\alpha$ values to fully characterize the distribution of consideration sets. This formulation allows us to estimate from the same observed market share data using a MLE technique consistent with that employed for the MNL, RCL, and NML models. The CTC model we estimate does not, however, then reflect the level of generality and efficiency available in the applications we review above. This does not affect our main purpose, to demonstrate the impact on design of non-compensatory behavior. 

\subsection{Connecting Nested Logit with Consideration}
\label{SUBSEC:NestedLogitCTC}

A few comments regarding the connection between the NML and CTC models are required, motivated by the similarity in the choice probabilities in the CTC and NML models. If the consideration sets used in the population are disjoint, then Eqn. (\ref{EQN:CTCjointProb}) describes the choice probabilities in a single-level nested Logit model whose nests are given by the consideration sets. However a NML would use the specific parameterization of $P_{\set{C}_r}$ given in Eqn. (\ref{EQN:Nest Probability}). It is easy to see that any true value of $P_{\set{C}_r}$ can be recovered in this parameterization by taking the nesting parameter $\lambda_r$ to be $1$ and choosing the right value of the coefficients for attributes that are constant over the consideration set (e.g., body style). Swait \cite{Swait2001a} has linked the generalized nested logit model to construct a general consideration set explosion model. Similarly, it is also easy to show that a single-level cross-nested Logit model \cite{Bierlaire2006} can realize choice processes as described in Eqn. (\ref{EQN:CTCjointProb}).

Though these models are the mathematically similar their interpretations differ, which drives a non-trivial difference in formalization. The NML model pictures rational, compensatory individuals that might use any consideration set and choose any product, and models consideration set frequency as a function of the expected maximum utility of choosing from a given consideration set \cite{Train2009}. While a NML can {\em recover} CTC behavior choosing the right parameters, it does not necessarily result in the same predictions as designs change because nest selection is a function of in-nest utilities. In contrast, the CTC model views individuals as drawn from a population with heterogeneous screening rules, and decouples consideration set frequency from utility. When consideration set occurrence is, completely or partially, independent of compensatory utilities, this distinction is meaningful.


\section{DESIGN OPTIMIZATION}
\label{SEC:Design_optimization}

This section defines a single firm's optimal vehicle design problem matching the stylized market model discussed above. The firm's objective is to maximize the expected profit of its vehicle portfolio by deciding the number of vehicles $J_f$ and choosing the body style, fuel economy, acceleration, and price for each vehicle. We allow firms to offer multiple vehicles of the same body style, as this is observed in real auto markets. 


\subsection{Engineering Model}
\label{SUBSEC:EngineeringModel}

Each vehicle is described by its 0-60 acceleration time ($a$, in s), fuel economy ($e$, in mpg), weight ($w$, $10^3$ lbs), body style ($b \in \{1,\dotsc,B\}$), ``technology content'' ($t$, unitless), and price $(p, \$10^4 )$. In the original model by \cite{Whitefoot2013}, $t$ is a continuous proxy for efficiency improvement through adoption of discrete technology content; this efficiency improvement can be directed towards either fuel economy or acceleration performance. To accomplish this, and to represent a physical connection between acceleration and fuel economy, $e,a,w$ and $t$ are related by a function $g_b(e,a,t)$ as given in Eqn. (\ref{design_const}):
\begin{equation}
	\begin{aligned}
	\label{design_const}
		g_b(e,a,t) &= \frac{1000}{e - 3.46} - \beta_{const}^g(b)  - \beta_{a}^g(b) \exp\{-a\} - \beta_{t}^g(b)t \\
				&\quad\quad\quad\quad - \beta_{at}^g(b)a^{2}t - \beta_{w}^g(b)w - \beta_{wa}^g(b)wa
	\end{aligned}
\end{equation}
Eqn. (\ref{design_const}) can be written as an equality constraint $g_b(e,a,t) = 0$ on acceleration and fuel economy decisions. Unit costs are also a function of design variables expressed by the following function:
\begin{equation}
	\begin{aligned}
	\label{cost_const}
		c_b(e,a) &= \beta_{const}^c(b) + \beta_{a}^c(b)\exp\{-a\} + \beta_{t}^c(b)t \\
				&\quad\quad\quad\quad + \beta_{w}^c(b)w + \beta_{wa}^c(b)wa
	\end{aligned}
\end{equation}
the body style specific coefficients in these models were estimated using detailed engineering simulations from AVL Cruise in conjunction with confidential technology production cost data provided to NHTSA by automakers in advance of the $2012-2016$ fuel economy rule making \cite{Whitefoot2013}. Table \ref{TBL:coef_fuelcons} and \ref{TBL:coef_cost} summarizes body style specific coefficient values. In our case study we assume that vehicle weight and technology content for each vehicle are fixed, and thus do not include these as arguments in $g_b$ or $c_b$. We use the average curb weights listed in Table \ref{TBL:weight}, based on 2005 model year vehicle data as reported in \cite{Wenzel2010}, and technology content $t_b = 20$. 

\subsection{Formulation and Solution}
\label{SUBSEC:Method}

Given a portfolio with $J_f$ vehicles and body style vector $\vec{b}$ the optimal choices of fuel economy, acceleration, and price for each vehicle are those that solve
\begin{equation}
	\label{EQN:OptimalDesign-FixD}
	\begin{aligned}
		\mxm &\quad \pi(\vec{p},\vec{e},\vec{a} | J_f, \vec{b}) \\
		\wrt 	&\quad \forall j, p_j \geq 0 \; \\
			&\quad L_{e,b(j)} \leq e_j \leq U_{e,b(j)}, \\
		         &\quad L_{a,b(j)} \leq a_j \leq U_{a,b(j)},  \\
		\sto 	&\quad g_{b(j)}( e_j , a_j ) = 0 \;\; \forall j
	\end{aligned}
\end{equation}
where expected profits are
\begin{equation}
	\pi(\vec{p},\vec{e},\vec{a} | J_f, \vec{b}) 
			= \sum_{j=1}^{J_f} P_j(\vec{p},\vec{e},\vec{a},\vec{b}) ( p_j - c_j(e_j,a_j) )
\end{equation}
and $(L_{e,b},U_{e,b}),(L_{a,b},U_{a,b})$ are body-style specific lower and upper bounds on fuel economy and acceleration. Note that we are not specific about what probability model we use. Eqn. (\ref{EQN:OptimalDesign-FixD}) is smooth for any of the models, because choosing prices, fuel economy, and acceleration does not affect screening in the CTC or, similarly, the nesting structure in NML. 

The optimal number of vehicles, body styles, and associated designs and prices can be obtained by solving
\begin{equation}
	\label{EQN:FixDB}
	\begin{aligned}
		\mxm &\quad \pi^*(J_f, \vec{b}) \\
		\wrt 	&\quad J_f \in \{1,\dotsc,B\}, \\
			&\quad b_j \in \{1,\dotsc,B\} \text{ for all } j = 1, ..., J_f\\
	\end{aligned}
\end{equation}
where $\pi^*(J_f, \vec{b})$ is the optimal value of Eqn. (\ref{EQN:OptimalDesign-FixD}) for given $J_f$ and $\vec{b}$. Note that we allow for multiple vehicles with the same body style. Because enumerating all all the feasible choices of body styles $\vec{b}$ for $1,\dotsc,B$ vehicles is computationally prohibitive, we use a Genetic Algorithm (GA) to solve Eqn. (\ref{EQN:FixDB}); our scheme is described in Appendix \ref{APP:GAscheme}. 


\section{RESULTS}
\label{SEC:Results}

This section presents performance results pertaining to choice model accuracy or predictive power, design ``error'', and profitability potential. To investigate how the amount of market information influences performance, we performed the simulation experiment with $M = 10, 25, 50, 100, 200, 500,$ and $1000$ markets. For each $M$ we draw 20 {\em separate} sets of $J_m = 5$ profiles and $N_m = 100$ choice observations, estimate MNL, RCL, NML, and CTC models, and then use these models to design product portfolios obtaining 20 {\em separate} sets of model estimates and designs. Sampling different sets of share data for a given market size allows us to gauge the effect of sampling variance in the data on model predictions, design outcomes, and design value, while examining different numbers of markets allows us to assess the asymptotic properties of the estimated models and their associated designs. The MLE and design optimization routines were programmed in C language, and nonlinear programs involved in estimation and design optimization were solved with the sequential quadratic programming (SQP) solver SNOPT (version 7) \cite{Gill2005}. All computations were undertaken on a single Mac Pro tower with 2, quad-core 2.26GHz processors and 32GB of RAM running OS X (10.6.8).


\subsection{Predictive Power}

The predictive power of the estimated models is validated on a new data set that consists of $M^\prime$ markets, where each market $m^\prime = 1, ..., M^\prime$ has a set of vehicles $\set{J}_{m^\prime}$. Kullback-Leibler Divergence \cite{Kullback1951}, 
\begin{equation}
	\label{EQN:KLD}
KLD = \left( \frac{1}{M^\prime} \right) \sum_{m^\prime=1}^{M^\prime} \sum_{j\in \set{J}_{m'} }  P^T_{j,m^\prime} \log \left( \frac{P^T_{j,m^\prime}}{P_{j,m^\prime}} \right), 
\end{equation}
captures how close the predicted choice probability distribution $P$ is to the actual choice probability distribution $P^T$ in the validation set. Predictive share errors are evaluated via Eqn. (\ref{EQN:KLD}) using 1,000 markets of validation data different from the estimation data, but drawn using the same approach. 

\begin{figure}
	\begin{center}
	\includegraphics{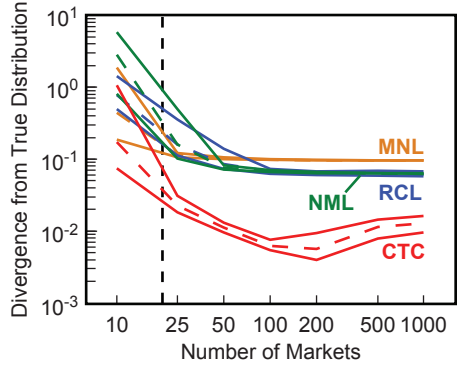}
	\caption{Kullback-Leibler Divergence (KLD) of predicted choice probability distribution from true behavior. Solid lines represent the range of observed values over 20 separate data sets while the dashed line represents the average value.}
	\label{Fig:DPvsM}
	\end{center}
\end{figure}

Fig. \ref{Fig:DPvsM} plots the divergence between predictions for estimated models and the true behavior against the number of markets used to train the models. Increasing the amount of market data available for estimation reduces both expected prediction error and the variance of the error. Increasing the amount of data, however, does not result in traditional compensatory models that match the predictive power of the CTC model. For example, the divergence of the three traditional models' predictions observing 1000 markets is larger than the CTC prediction observing only 10 markets. When observing more than 50 markets, the predictive power of RCL and NML models is generally between those of MNL and CTC with RCL predictions appearing to be slightly closer to the true behavior. However, when observing fewer than 10 markets the MNL model outperforms RCL and NML models. We believe designers should be particularly interested in performance when estimating models with relatively small amount of market data because real revealed preference market research often uses a very limited number of markets for estimation. For example, econometric new vehicle market models most often use fewer than 20 markets (marked with vertical line in Fig. \ref{Fig:DPvsM}) \cite{Berry1995, Goldberg1995, Golob1997, Train2009}. Our market simulation setting is not strictly comparable to these studies because of a difference in the number of vehicle-observations in each market, the complexity of real vehicle profiles, and the detail often given by population demographics. But these results suggest caution given the small number of markets usually used for model estimation in practice. 

\subsection{Decision Bias and Variance}

We first define a ``design error'' metric to quantify how different portfolios chosen using an estimated model are from portfolios that would be chosen for the true behavior (perfect information). Comparing product portfolios is a complicated task, and we do not suggest we have a uniquely good metric for comparison. Essentially, our metric compares the relative difference in specific vehicle attributes $-$ excluding price to focus on engineering aspects of strategy $-$ for the same styles of vehicles and number of vehicles with different body styles. The specific numerical values of ``design error'' for any given choice model are less important than comparisons across the different choice model types we explore.

Suppose a portfolio has $J_f$ vehicles, each with body style $b_j$ and design $\vec{x}_j = (e_j, a_j)$. Denote the body style combinations of a portfolio as $(n_1, n_2, ..., n_B)$ where $n_b$ is the number of vehicles in the portfolio that have body style $b$. We refer to the \textit{ideal} portfolio as an estimate of the globally optimal portfolio with the true behavior and denote the ideal portfolio with superscript ``*''; multiplicity of ideal portfolios is addressed below. Our design error metric is
\begin{equation}
\label{EQN: design_error}
\begin{aligned}
	d = \frac{1}{2} \left( \sum_{b=1}^B N_b + \max \{ H^+ , H^- \} \right)
\end{aligned}
\end{equation}
where
\begin{align}
	&N_b
		= \left\{ \begin{aligned} 
			&| n_b - n_b^* |/n_b^* &&\quad\text{if } n_b^* > 0  \\
			&\quad\quad n_b  &&\quad\text{if } n_b^* = 0 
		\end{aligned}  \right. 
		\label{EQN:DErrBodyStyle} \\ 
	&H^+= \max_{j:n^*_{b(j)} > 0}  \Bigg\{ \min_{k:b^*(k)=b(j)}  \Bigg\{ d_w (\vec{x}_j , \vec{x}_k^* ) \Bigg\} \Bigg\}
		\label{EQN:HausdorffPlus} \\
	&H^- = \max_{j:n_{b(k)} > 0}  \Bigg\{ \min_{j:b(j)=b^*(k)}  \Bigg\{ d_w (\vec{x}_j , \vec{x}^*_k ) \Bigg\} \Bigg\}
		\label{EQN:HausdorffMinus} \\
	&d_w(\vec{x} , \vec{x}^* ) = \frac{1}{2} \left( \frac{|e - e^*|}{e^*} + \frac{|a - a^*|}{a^*} \right)
		\label{EQN:SimpleDesignError}
\end{align}
The first term in Eqn. (\ref{EQN: design_error}), Eqn. (\ref{EQN:DErrBodyStyle}), captures differences in body style combinations by penalizing differences in the number of vehicles offered with each body style. The second term in Eqn. (\ref{EQN: design_error}), composed of Eqns. (\ref{EQN:HausdorffPlus}-\ref{EQN:SimpleDesignError}), is a Hausdorff distance \cite{Rockafellar05} comparing {\em sets} of vehicles with the same body styles using the relative error metric in Eqn. (\ref{EQN:SimpleDesignError}). This portion of the metric is zero so long as the sets of vehicles offered are equivalent, even if offered in different multiplicities. If nonzero, this portion gives the relative error in the attributes of any vehicle offered when that vehicle shares a body style with a body style offered in the ideally optimal portfolio. Note that this distance measure gives an error only in engineering decisions, while pricing is obviously important to profitable product design. However it is plausible that ``incorrect'' prices could be corrected relatively quickly in the marketplace after offering a particular set of products, while errors in engineering features cannot be. Section \ref{SUBSEC:PricingBias} explores this in more detail. See \cite{Long14} for a generalization of Eqn. (\ref{EQN: design_error})-(\ref{EQN:SimpleDesignError}) that accounts for prices. 

\begin{figure}
	\begin{center}
	\includegraphics[width=3.25in]{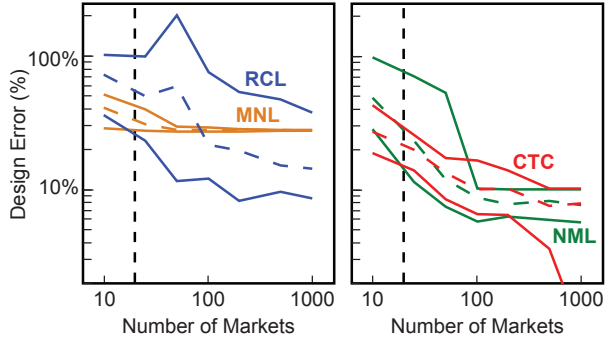}
	\caption{Design error, measured as in Eqn. (\ref{EQN: design_error}), for all models over all numbers of markets observed. Solid lines represent the range of observed values over 20 separate data sets while the dashed line represents the average value.}
	\label{Fig:DesignError_all}
	\end{center}
\end{figure}

Fig. \ref{Fig:DesignError_all} plots design error for optimal decisions based on the MNL, RCL, NML, and CTC choice models against number of markets observed. Two features are of interest: design {\em bias} refers to the difference between mean model-optimal designs and ideal optimal designs; design {\em variance} refers to the spread of designs that might be made given different observed markets used to estimate the choice models. With fewer than 25 markets the CTC model has the lowest design bias, consistent with the performance this model showed in predictive power. As the amount of market data grows, designs under the CTC model appear to be converging to ideal designs. Designs chosen using a NML model compare well to CTC designs when observing 50 or more markets. Designs chosen using a RCL model have the largest variation among the models, and this variation cannot be overcome by increasing the number of markets observed. Unlike RCL, NML and CTC models, using a MNL model suggests identical vehicles with the same body style should be produced; this is reflected as the high design error compared to the ideal design in which there is wide diversity among body styles.

In computing design error we presume that a computed ideal portfolio is a good representation of portfolios required to achieve global optimality relative to the true behavior. If there are distinct locally optimal portfolios that achieve nearly globally optimal profits a different metric would be required. Similarly, if profits were ``flat'' near the ideal portfolio design error loses meaning. In the next section we discuss decision profitability, which is the ultimate metric of portfolio performance.

However, an important issue in our use of this metric pertains to whether our ``ideal'' optimal designs are a good representation of the global optimum. Given suitable evolution rules, GAs can explore the entire design space and thus global optimality is guaranteed for discrete variable problems, but only because the design space can be enumerated; global optimality cannot be guaranteed in finite time, with probability one, for continuous variable problems. In our case, we applied a GA to a relatively small, $B(B+1) = 90$ binary variable problem, evolving populations for 50-100 generations (until the best and average fitness values converged), and used 50 trial-multistart. Given the small size of design space, this process is very likely to obtain good solutions. We found a high degree of consistency in both the body type combinations obtained by the GA and the design and price for each body type found by the NLP; in terms of our own design metric, the optimal solutions found by multistart differed by less than $10^{-6}$. Thus we have high confidence that our computed ideal optimal designs are as good as could be practically obtained. 


\subsection{Decision Profitability}

A decision that differs from an ideal decision is not necessarily un-profitable. The \textit{true} profit of a product portfolio is its expected profits computed under the true behavior, rather than the estimated model. We compute true profit $\Pi$ as
\begin{equation}
\label{EQN:Synthetic Profit}
	\Pi = \sum_{j=1}^{J_f} P_j^T(\vec{e},\vec{a},\vec{p},\bsym{\triangle}) ( p_j - c_j) 
\end{equation}
computing market share $P_j^T$ by computing the choice probability in the true behavior described in Section \ref{SUBSEC:SyntheticBehavior} instead of the sampling procedure described in Section \ref{SUBSEC:MarketShareSimulation}. Here the Monte-Carlo sampling size used to approximate the compensatory stage random coefficients model was $I = 100,000$. We do not include competitive firms' vehicles in this profit validation in order to be consistent with the design optimization problems, which do not include competition. Note that true profit for any model-optimal portfolio should be less than the true profit given by the optimal portfolio under the true behavior. We refer the the profit gained by optimal decisions under the true model as the \textit{ideal} profit. 

\begin{figure*}
	\begin{center}
	\begin{minipage}[c]{3.25in}
	\includegraphics[width=3.25in]{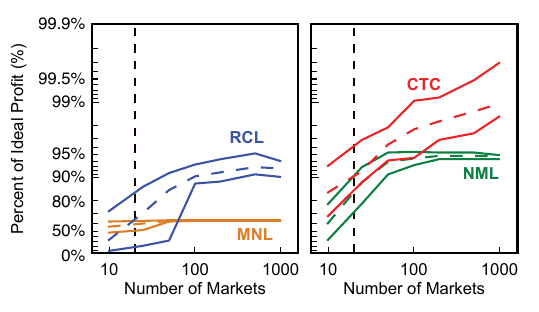}
	\caption{Percent of ideal profits obtained by designs and prices under true behavior recovery when choosing designs and prices with estimated models. Note the $\log_{10}$ scale y-axis focuses on differences from 100\%. Solid lines represent the range of observed values over 20 separate data sets while the dashed line represents the average value.}
	\label{Fig:ProfitError}
	\end{minipage}
	\hfill
	\begin{minipage}[c]{3.25in}
	\includegraphics[width=3.25in]{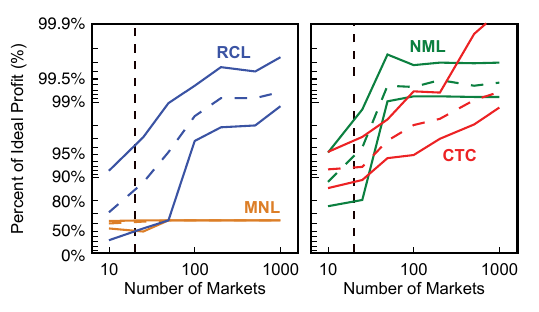}
	\caption{Percent of ideal profits obtained by designs chosen using estimated models, but optimizing prices for these designs with knowledge of the true choice behavior. Note the $\log_{10}$ scale y-axis focuses on differences from 100\%. Solid lines represent the range of observed values over 20 separate data sets while the dashed line represents the average value.}
	\label{Fig:PricingBias}
	\end{minipage}
	\end{center}
\end{figure*}

Fig. \ref{Fig:ProfitError} plots the percentage of ideal profits that can be achieved by choosing designs and prices using an estimated model. CTC designs and prices are, not surprisingly, best able to capture true profits. Even when observing only 10 markets, CTC designs and prices can be expected to achieved at least 90\% of the ideal profits with the worst designs achieving around 70\% of the ideal profits. MNL designs and prices can be expected to obtain only 60\% of the ideal profit due to a single body-style portfolio that lacks diversity. Estimating a model with limited amount of market data appears to affect the profitability of the RCL model designs the most out of all the models, consistent with our observations regarding design error variance. Even observing up to 50 markets it is {\em possible} for RCL designs and prices to recover less than 40\% of the ideal profits, depending on sampling variance in the market data. More data results in RCL designs and prices within 90-95\% of the ideal profits. Observing less than 25 markets NML designs capture approximately 20\% less true profit than the CTC designs and also shows relatively high variation in true profits (facts obscured by the log$_10$-scale axis in Fig. \ref{Fig:ProfitError}). However like RCL, NML designs and prices can ultimately capture 90-95\% of the true ideal profits when estimating the model with enough data. 

\subsection{Pricing-On-Offering}
\label{SUBSEC:PricingBias}

An additional test assesses the degree to which a model suggests unprofitable decisions simply because of a poor representation of preferences over prices. Prices can, in principle, be changed up until the point-of-sale while design decisions must often be fixed far in advance of sale. Thus it is reasonable to consider a case where firms learn more about preferences when they offer the portfolio designed and exercise price flexibility to maximize profits. Suppose that the MNL, RCL, NML, and CTC models inform the design of the product portfolio but that prices can be changed even the product is offered (as in, e.g., \cite{Morrow2014a}). How much more profits could the firm recover by using the true choice behavior in order to set optimal prices, for fixed designs? While the firm is not likely to actually know the true behavior, this value represents an upper bound on profitability of design decisions made using an estimated model when prices are flexible and determined when offering the portfolio. 

Fig. \ref{Fig:PricingBias} plots percent of ideal profits obtained using the vehicle portfolios suggested by the estimated models, but offered at prices determined by the optimizing profits for that portfolio under the true behavior. From this perspective the RCL, NML, and CTC models each have the potential to suggest nearly equivalently profitable design decisions. RCL and NML, in particular, can suggest much more profitable portfolios if we admit pricing flexibility than if we donÕt, and thus RCL and NML capture pricing preferences more weakly than does the CTC model. Moreover, for intermediate numbers of markets (25, 50, 100, and 200), the NML model appears to suggest the most profitable portfolios by a small margin (less than 2.7\%) that is exaggerated by the log$_{10}$ axis scaling. Finally, even the best possible pricing strategy cannot increase the true profitability of the single body-style portfolio designed under the MNL model. 


\section{DISCUSSION AND LIMITATIONS}
\label{SEC:Discussion}

There are several observations for designers to take away from this exploratory simulation study. 

First, conventional compensatory models can reasonably support profitable design decisions even when the population exhibits non-compensatory behavior {\em with enough data}. Designs based on estimated RCL and NML models were capable of obtaining above 90\% of the ideal profits (Fig. \ref{Fig:ProfitError}); however this required roughly twice the amount of market data (50 markets) that might typically be available (20 markets) judging from the vehicle modeling literature. The RCL and NML models could suggest designs that obtain almost 100\% when an ideal pricing strategy is followed (Fig. \ref{Fig:PricingBias}), but this would require learning preferences {\em exactly} when actually offering the vehicles designed (Fig. \ref{Fig:PricingBias}). This is practically impossible but does suggest that a significant portion of the ``error'' made with conventional models pertains to pricing bias, not design bias. However designers should be aware of the possible side effects of different compensatory model structures: The MNL model suggests portfolios with identical body styles; the RCL model, estimated on limited amounts of aggregate market share data, is highly sensitive to sample error leading to large variations in optimal designs; and the NML model, while it might capture optimal designs very well if the nesting structure reflects consideration, suggests biased pricing decisions and thus cannot present accurate forecasts of design profitability. Designers also need to take into account the amount of information available to train their model when they decide what model to use. According to our simulation using the MNL might be more {\em reliably} profitable than using the RCL and NML models if market data are very limited, because noise in the data induces greater variance in designs suggested by RCL and NML models. 

Second, modeling the heterogeneity in the screening rules may capture more value to design than modeling heterogeneity in the compensatory stage. This is most directly observed by comparing the CTC and RCL models. The RCL model ignores screening stage heterogeneity, and achieved only 30\% of the ideal profit (on average) with 10 markets while displaying an unacceptably large sensitivity to sample variance with limited training data. The CTC model with only 10 marketsÕ of training data gives a firm expected profit that is at least 80\% as much as what they could get with perfect knowledge. Recall that the CTC model is mis-specified, in that it ignores compensatory stage heterogeneity. Other evidence comes from the NML. NML and CTC are similar in a two-stage modeling structure. In effect, our NML is a close approximation to a CTC model assuming that individuals consider one, and only one, vehicle body style. Decisions made based on the NML are most often more profitably than those made with the RCL model for all amounts of training data. This observation may be driven by limited amount of heterogeneity in our assumed true behavior (see Table \ref{TAB:TrueCoeffs}), suggesting further research is required.

Note also that the CTC model has the potential to be seriously overfit. For example, the CTC model in with M = 50 has more than twice as many parameters (515) as observations (250), but is still the most predictively accurate model (Fig. \ref{Fig:DPvsM}, right) and results in the most profitable decisions (Fig. \ref{Fig:ProfitError}, right). Conventional wisdom would suggest that at least as many observations as parameters are required for a valid model; i.e., the CTC model requires, at a minimum, M ³ 103 markets with 5 vehicles per market. We believe that the stability of estimated model predictions is more important than the ratio of parameters to observations; Fig. \ref{Fig:DPvsM} shows that the predictive power of the estimated CTC model is as good as it can get with as few as 50 markets. However, overfitting effects may result in the difference between NML and CTC model performance when pricing on offering (Fig. \ref{Fig:PricingBias}): our CTC model presumes that all $511$ screening rules may be in use by the population generating the data, while the NML approximates a CTC model with only $B = 9$ screening rules. Slightly better performance with the simpler model suggests some overfitting may be occurring, although any such overfitting could be easily corrected by restricting the number of nonzero $\alpha$ coefficients in the CTC estimation. 

\begin{figure*}
	\begin{center}
	\includegraphics{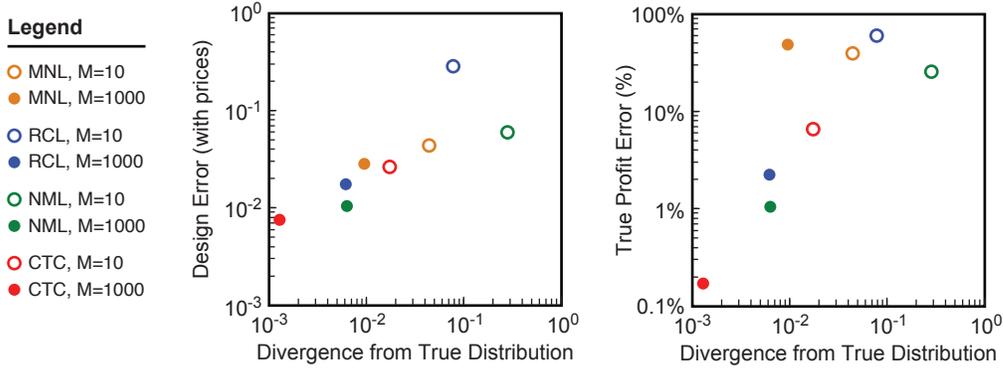}
	\caption{Expected design error (left) and expected Profit Error (right) versus Kullback-Leibler divergence of choice prediction. Completely recovery of the ideal profit yields 0\% error while zero profit yields 100\% error.}
	\label{Fig:DEvsKLD}
	\end{center}
\end{figure*}

Third, assuming that better predictive power indicates better design decisions is reasonable but not necessarily true. Pearson correlation coefficients are positive but weak: 0.62 between predictive power (divergence) and design error, 0.73 between predictive power and profit error (measured as error, not percent of ideal profits recovered), and 0.73 between design and profit errors. Fig. \ref{Fig:DEvsKLD} scatters the average design error and average true profitability versus the average Kullback-Leibler divergence of four models estimated under two market information conditions: 10 and 1,000 markets; Here, the average is  taken over different data sets with the same number of markets. While there is a general trend that lower divergence (better model predictive power) is consistent with lower design error, deviation from this trend is also observed. For example, NML has, on average, worse choice predictions but better designs than RCL for both 10 and 1,000 markets worth of data. Lower model divergence also generally indicates less loss of profit. However there are exceptions, such as the comparison between the NML and the RCL. Note also the difference in scales: the MNL model does not appear to predict that much worse than the NML or RCL models while suggesting designs that capture almost no profit relative to NML or RCL. 

These results shows that the true profitability of designs made using traditional models cannot be judged from predictive power alone. While further investigation of the relationship between predictive accuracy and decision value across a range of design problems and market conditions is needed, it is clear that choice models with structure representative of the underlying choice process are better for design even if they may not show significant benefits from the perspective of modeling choice alone.

Important limitations of our study are as follows. 

Our study has focused on the value of incorporating {\em prior} knowledge on screening without demonstrating the process needed to obtain that knowledge. Nor has our study examined the consequences of misspecified prior knowledge. The CTC model in this study is able to estimate the distribution of the possible consideration sets from choice data given that the attributes involved in the screening process are known and limited. Our presumed behavior$-$screening over body style$-$is a reasonable prior for the case study and is represented in some form in every model we tested. Aggregate share data is insufficient to infer what attributes and screens are involved in the consideration stage. We are currently mirroring this simulation study within the context of survey design for both choice-based conjoint and consideration-based questions \cite{Dzyabura2011} in which screening rules can be statistically inferred. Subjective beliefs, however, often inform choice model construction; they underlie decisions about what utility function to use and what distribution the error term takes (including heterogeneous preferences and nesting structures). While we must assume that misspecification of screening rules would impact design outcomes, this is a generic problem for choice modeling. 

The design problem in our case study is also simplistic. The engineering model merely focuses on a body style specific fuel consumption-acceleration relationship and cost function that depends only on fuel economy and acceleration. The screening rules we used were independent of continuous features such as price and fuel economy; in contrast, the study from which we drew screening rules estimated rules over a body style, brand, fuel economy, price, quality, safety, power, and powertrain \cite{Dzyabura2011}. Future studies should include more engineering model as well as complexities in screening. 

Bayesian methods \cite{Rossi2006} for model estimation were also not used in this study. An important fact is that Bayesian methods provide an alternative path to estimate the parameters of a choice model, not fundamentally different models. Theoretically speaking, maximum likelihood and Bayesian estimators are often similar; in particular the posterior mean of a Bayesian estimator is asymptotically equivalent to the maximum likelihood estimator \cite{Train2009}. Empirically Bayesian estimators have been reported to have better fit small-sample data but have predictive power and parameter recovery similar to maximum likelihood estimation \cite{Andrews2002}. In the context of our study we might then expect Bayesian estimators to achieve larger-sample performance with fewer data, but not to qualitatively change the comparison between conventional compensatory models and non-compensatory models when consumers consider. 


\section{CONCLUSIONS}
\label{SEC:Conclusion}

This paper explores the impact of consideration behavior on optimal design for market systems models. We presented a simulation study of vehicle portfolio design for a population with heterogeneous screening over body style and heterogenous compensatory evaluations after screening. With synthetically generated aggregate marketshare data we estimate multinomial Logit, random coefficient Logit, nested multinomial logit, and consider-then-choose logit models. All four models contain some representation of screening, and all are misspecified in at least one dimension of the true behavior. We use the estimated models to optimize designs for a single model and compare model performance in terms of predictive power, design error, and profitability. We find that capturing heterogeneous consideration, when it exists, is more important than capturing heterogeneous tradeoffs. This can be accomplished with consider-then-choose Logit, but also with the right nested Logit model. Decisions made using Logit models are simplistic, suggesting portfolios with a single body style, and decisions made using random coefficients Logit models are noisy; with limited amounts of data, Logit models may often lead to more profitable decisions than random coefficients Logit. We also observe that higher model predictive power generally does imply a more profitable design decision, but that there are cases where poorer predictors can yield higher profits. 


\bibliographystyle{asmems4}

\end{document}